\pdfoutput=1
\documentclass[12pt,letterpaper,UTF8,superscriptaddress,nofootinbib]{revtex4-2}
\usepackage{graphicx}
\usepackage{multirow}
\usepackage{amsmath,amssymb,amsfonts,amsbsy}
\usepackage{mathrsfs}
\usepackage{booktabs}
\usepackage{etoolbox}
\usepackage{csquotes}
\usepackage{xcolor}
\usepackage{longtable}
\usepackage{url}
\usepackage{natbib}
\usepackage{wrapfig}
\usepackage{amsthm}
\usepackage{fancyhdr}
\usepackage{empheq}
\usepackage{bm}
\usepackage{ulem}
\usepackage{cancel}

\usepackage{epstopdf}
\usepackage[super]{nth}
\usepackage{tabularx}
\usepackage[T1]{fontenc}
\usepackage[utf8]{inputenc}
\usepackage[activate={true,nocompatibility},final,tracking=true,kerning=true,spacing=true]{microtype}
\usepackage[colorlinks = true, %Colours links instead of stupid boxes
            urlcolor   = blue, %Colour for external hyperlinks
            linkcolor  = blue, %Colour of internal links
            citecolor  = blue,
            linktoc    = all,
            bookmarks  = true,
            bookmarksnumbered = true,
            bookmarksdepth = 4,
            hyperindex,breaklinks]{hyperref}

\begin{document}
\title{RPEM: Randomized Monte Carlo Parametric Expectation Maximization Algorithm}
\author{Rong Chen}
\email{rongchen@chla.usc.edu}
\affiliation{Laboratory of Applied Pharmacokinetics and Bioinformatics, Children's Hospital Los Angeles, Keck School of Medicine, University of Southern California, Los Angeles, CA 90027, USA}
\affiliation{Department of Physics,
Arizona State University, Tempe, Arizona 85287, USA}
\author{Alan Schumitzky}
\email{schumitzky@gmail.com}
\affiliation{Laboratory of Applied Pharmacokinetics and Bioinformatics, Children's Hospital Los Angeles, Keck School of Medicine, University of Southern California, Los Angeles, CA 90027, USA}
\affiliation{Department of Mathematics, University of Southern California, Los Angeles, CA 90089, USA}
\author{Alona Kryshchenko}
\email{alona.kryshchenko@csuci.edu}
\affiliation{Department of Mathematics, California State University Channel Islands, University Dr, Camarillo, CA 93012, USA}
\author{Romain Garreau}
\email{romain.garreau@chu-lyon.fr}
\affiliation{UMR CNRS 5558, Laboratoire de Biométrie et Biologie Evolutive, Université de Lyon, Université Lyon 1, Villeurbanne, France  }
\author{Julian Otalvaro}
\email{jotalvaro@chla.usc.edu}
\author{Walter Yamada}
\email{wyamada@chla.usc.edu}
\affiliation{Laboratory of Applied Pharmacokinetics and Bioinformatics, Children's Hospital Los Angeles, Keck School of Medicine, University of Southern California, Los Angeles, CA 90027, USA}
\author{Michael N. Neely}
\email{mneely@chla.usc.edu}
\affiliation{Laboratory of Applied Pharmacokinetics and Bioinformatics, Children's Hospital Los Angeles, Keck School of Medicine, University of Southern California, Los Angeles, CA 90027, USA}
\affiliation{Pediatric Infectious Diseases, Children's Hospital Los Angeles, Keck School of Medicine, University of Southern California, Los Angeles, CA 90027, USA}
\date{\today}
\begin{abstract}
Inspired from quantum Monte Carlo, by using unbiased estimators all the time and sampling discrete and continuous variables at the same time using Metropolis algorithm,
we present a novel, fast, and accurate high performance Monte Carlo Parametric Expectation Maximization (MCPEM) algorithm.
We named it Randomized Parametric Expectation Maximization (RPEM).
In particular, we compared RPEM with Monolix's SAEM and Certara's QRPEM for a realistic two-compartment Voriconazole model with ordinary differential equations (ODEs) and using simulated data.
We show that RPEM is 3 to 4 times faster than SAEM and QRPEM for the Voriconazole model,
and more accurate than them in reconstructing the population parameters.
\end{abstract}

\maketitle
%\tableofcontents % turn it off can significantly decrease compiling time.
\allowdisplaybreaks
\section{Introduction}
The expectation maximization (EM) method \cite{Dempster1977a} is a widely used powerful algorithm in machine learning, including population modeling of pharmacokinetics (PK) and pharmacodynamic (PD) systems.
EM treats data in terms of `complete' and `missing'. By using Bayes Theorem, the missing data are integrated out, and the parameters of the parametric model are learned automatically through iterations between expectation steps (E-step) and maximization steps (M-step).
An EM algorithm based, exact maximum likelihood solution to the parametric population modeling problem was proposed by Alan Schumitzky \cite{schumitzky1995algorithms} in 1995 and fully implemented by Stephen Walker \cite{walker1996algorithm} in 1996 for non-mixture models.
For mixture models,  the corresponding formulas were derived in \cite{wang2007nonlinear}.

\textcolor{black}{
In general, the EM algorithm needs to address two important problems.
One is from the theoretical aspect, namely the convergence problem.
EM algorithms converge to a stationary point \cite{Wu1983a,Tseng2004a,wang2007nonlinear} of the likelihood function from the given initial conditions.
However, a stationary point can be a local maximum, a minimum, or a saddle point, and therefore it may not be the global maximum of the likelihood function.
A standard method to increase the probability of finding the global maximum is to repeatedly run the algorithm, initializing each run with different conditions. This of course is computationally intense, so developing efficient methods to avoid converging on solutions which are not truly maximally likely is an important task.}

\textcolor{black}{
We focus in the paper on the second problem which faces EM algorithms from the computational aspect, namely how to efficiently and accurately evaluate all the integrals in the E-step and the M-step. This is important because the integrals have to be evaluated at each iteration.
If the integrals can be evaluated efficiently,
we can reach the stationary point rapidly,
which will enable us to search from more initial conditions per unit time, and therefore also help addressing the first problem of convergence described above.
Parametric EM methods typically use Monte Carlo algorithms to evaluate the integrals, and such methods are called Monte Carlo parametric EM methods (MCPEM).
Therefore, fast, accurate and robust Monte Carlo algorithms are always desirable for MCPEM.
}

\textcolor{black}{
There are successful and widely used Monte Carlo engines implemented in software tools commonly used for PK/PD population modeling, such as those used in ADAPT from the University of Southern California \cite{ADAPT},  S-ADAPT \cite{S-ADAPT,sadapt}, ICON's NONMEM program \cite{NONMEM,NONMEMwiki}, Quasi-random parametric EM (QRPEM) in Certara's Phoenix software and RsNLME (R Speaks Non-Linear Mixed Effects Modeling) \cite{QRPEM,RsNLME}, and Stochastic approximation EM (SAEM) \cite{SAEM1999,kuhn2004coupling,KUHN20051020,lavielle2014improved} in Lixoft's Monolix software.
As far as we know, each of these MCPEM algorithms strictly sums over the Monte Carlo integrals for all the $n$ subjects, and, when applicable, for all the possible $K$  components of a mixture model to describe the joint model parameter value probability distributions as sums of normal distributions.
}

Inspired from quantum Monte Carlo methods \cite{Lomnitz1981NPA, Ceperley95a,CarlsonRMP15a,CR2020PhD,chen2022path} which is one of the most precise, reliable and successful computational methods widely used in physics and chemistry,
in this work we approach the Monte Carlo integrals in a novel way by developing our randomized parametric expectation maximization (RPEM) algorithm.
It uses an unbiased Metropolis method to efficiently sample from the subjects and mixture models instead of strictly summing over all of them.
The unique Metropolis method in M-step allows us to efficiently get unbiased estimations for both analytic models \cite{wang2007nonlinear,wang2009population} and models with ODEs without taking many samples in E-step which is time consuming.
Therefore, as we will show, RPEM is fast without compromising accuracy.

The paper is organized as follows.
In Sec. \ref{secMethods}, we briefly introduce the two-stage nonlinear random effects mixture model \cite{wang2007nonlinear} we use and the EM algorithm. Then we describe the algorithm of RPEM.
In Sec. \ref{secResults}, we report the results of RPEM by using a concrete example of the two-stage nonlinear random effects two-mixture model \cite{wang2007nonlinear},
as well as a realistic Voriconazole model with ODEs.
We compare RPEM with SAEM and QRPEM and we show that RPEM is not only fast but also accurate to varied starting conditions.
We also report how RPEM scales on the supercomputer.
In Sec. \ref{secSummary}, we summarize the novelty, speed, and accuracy  of RPEM. Possible future directions are considered in closing.

\section{Methods}
\label{secMethods}

\subsection{Two-Stage Nonlinear Random Effects Mixture Model}

We use a two-stage nonlinear random effects model \cite{wang2007nonlinear} as an example.
At stage one,
given $\bm{\theta}_i$ which is the parameter vector describing the random effects ($\theta_i \in R^p$),
as well as $\bm{\beta}$ which describes the fixed effects ($\bm{\beta} \in R^q$),
the $m_i$-dimensional observation vector for the $i$th individual $\bm{Y}_i=(Y_{1 i}, ..., Y_{m_i i})^T$ is sampled from a Gaussian distribution such that,
\begin{equation}
\bm{Y}_i \vert \bm{\theta}_i, \bm{\beta} \sim N(\bm{h}_i(\bm{\theta}_i),\bm{G}_i(\bm{\theta}_i,\bm{\beta})),
~~ i=1,...,n
\label{stage1}
\end{equation}
where $n$ represents the number of subjects,
$\bm{h}_i(\bm{\theta}_i)$ is the function defining the PK/PD model, and
$\bm{G}_i(\bm{\theta}_i,\bm{\beta})$ is a positive definite covariance matrix ($\bm{G}_i \in R^{m_i \times m_i}$).
In this paper we consider the following important case \cite{wang2007nonlinear},
\begin{eqnarray}
\bm{G}_i(\bm{\theta}_i,\bm{\beta}) = \sigma^2 \bm{H}_i(\bm{\theta}_i),
\end{eqnarray}
where $\bm{H}_i(\bm{\theta}_i)$ is a known function and $\bm{\beta} = \sigma^2$ .

At stage two, each of the $n$ parameter vectors $\bm{\theta}_1$, ...,  $\bm{\theta}_n$ is sampled from Gaussian distributions with $K$ mixing components,
\begin{eqnarray}
\bm{\theta}_1, ..., \bm{\theta}_n \sim_{i.i.d} \sum\limits_{k=1}^{K} w_k N(\bm{\mu}_k,\bm{\Sigma}_k),
\label{stage2}
\end{eqnarray}
where nonnegative number $w_k$ (normalized by $\sum\limits_{k=1}^{K} w_k=1$) is the weight for the $k$th Gaussian distribution $N(\bm{\mu}_k,\bm{\Sigma}_k)$, $\bm{\mu}_k$ is the mean vector ($\bm{\mu}_k \in R^p$) and $\bm{\Sigma}_k$ is the positive definite covariance matrix ($\bm{\Sigma}_k \in R^{p \times p}$).

Given the observation data $\{ \bm{Y}_1, ..., \bm{Y}_n  \}$, we want to estimate $\phi$ which represents the collection of parameters $\{ \bm{\beta}, (w_k, \mu_k, \bm{\Sigma}_k), k=1,...,K  \}$ by maximizing the overall data likelihood $L(\phi)$ which is written as,
\begin{eqnarray}
L(\phi) = \prod_{i=1}^n  \sum\limits_{k=1}^{K}  \int p(\bm{Y}_i \vert \bm{\theta}_i, \bm{\beta}) w_k p(\bm{\theta}_i \vert \bm{\mu}_k, \bm{\Sigma}_k) d \bm{\theta}_i.
\label{L}
\end{eqnarray}
This is called the maximum likelihood estimate (MLE). The MLE of $\phi$ is defined as $\phi_{ML}$ such that $L(\phi_{ML})  \geq L(\phi) $ for all $\phi$ within the parameter space.

\subsection{EM Algorithm}

In the EM algorithm, we define complete data and missing data. In this model we define $\{\bm{\theta}_i, \bm{z}_i\}$ as missing data, where $\bm{z}_i$ is a $K$-dimensional vector whose $k$th component $z_i(k)$ is 1 or 0 depending on whether $\bm{\theta}_i$ belongs to the $k$th mixing in Eq.(\ref{stage2}) or not.
The complete data is defined as $\bm{Y}_c = \{ ( \bm{Y}_i,\bm{\theta}_i, \bm{z}_i ),i=1,...n  \}$.
The purpose of the EM algorithm is to start with $\phi^{(0)}$ and iterate from $\phi^{(r)}$ to $\phi^{(r+1)}$ at the $r$th iteration, continue the process until we find the desired parameters $\phi^{(r+1)}$ such that
$\phi^{(r+1)} = \underset{\phi}{\text{arg max}}\,Q(\phi, \phi^{(r)})$.
This process is guaranteed to converge to a stationary point of the likelihood \cite{wang2007nonlinear,Wu1983a,Tseng2004a},
and typically, a number of starting positions are suggested in an effort to ensure convergence to a global maximum.

In the E-step, the function $Q(\phi, \phi^{(r)})$ is defined as,
\begin{eqnarray}
Q(\phi, \phi^{(r)}) = E \left \{ \log L_c(\phi) \vert \bm{Y}, \phi^{(r)} \right \},
\label{Qdef}
\end{eqnarray}
where the complete data likelihood $\log L_c(\phi)$ is given by
\begin{eqnarray}
\log L_c(\phi) = \sum\limits_{i=1}^n \sum\limits_{k=1}^{K}
\log {p(\bm{Y}_i,\bm{\theta}_i \vert \sigma^2, \bm{\mu}_k, \bm{\Sigma}_k ) }^{z_i(k)}.
\end{eqnarray}

By using Bayes Theorem,
Eq.(\ref{Qdef}) can be concretely written as \cite{wang2007nonlinear},
\begin{eqnarray}
Q(\phi, \phi^{(r)}) =
\sum\limits_{i=1}^n \sum\limits_{k=1}^{K} \int g_{ik}(\bm{\theta}_i,\phi^{(r)}) \log p(\bm{Y}_i, \bm{\theta}_i \vert \sigma^2, \bm{\mu}_k, \bm{\Sigma}_k) d \bm{\theta}_i ,
\label{Qphiphir}
\end{eqnarray}
where
\begin{align}
g_{ik}(\bm{\theta}_i,\phi)
= \frac{w_k p(\bm{Y}_i\vert \sigma^2,\bm{\theta}_i)
p(\bm{\theta}_i \vert \bm{\mu}_k,\bm{\Sigma}_k )   }{ \displaystyle \sum\limits_{k=1}^{K} w_k \int p(\bm{Y}_i\vert \sigma^2,\bm{\theta}_i) p(\bm{\theta}_i \vert \bm{\mu}_k,\bm{\Sigma}_k ) d \bm{\theta}_i },
\label{gik}
\end{align}
and
\begin{align}
\log p(\bm{Y}_i, \bm{\theta}_i \vert \sigma^2, \bm{\mu}_k, \bm{\Sigma}_k) &=
C - \frac{m_i}{2} \log(\sigma^2) \nonumber \\
& - \frac{1}{2\sigma^2}(\bm{Y}_i-\bm{h}_i(\bm{\theta}_i))^T {\bm{H}_i(\bm{\theta}_i)}^{-1}(\bm{Y}_i-\bm{h}_i(\bm{\theta}_i)) \nonumber \\
& - \frac{1}{2}( \bm{\theta}_i - \bm{\mu}_k )^T \bm{\Sigma}_k^{-1} ( \bm{\theta}_i - \bm{\mu}_k )
-\frac{1}{2}\log\vert\bm{\Sigma}_k\vert,
\end{align}
for some constant $C$.
Note that the probability that the $i$th individual belongs to the $k$th mixing component can be defined as a function $\tau_i(k)$ such that
\begin{eqnarray}
\tau_i(k)=E\{ z_i(k)\vert \bm{Y},\phi \}
= \text{pr} \{ z_i(k)
= 1 \vert \bm{Y},\phi \}
= \int g_{ik} (\bm{\theta}_i,\phi) d \bm{\theta}_i .
\label{tauik}
\end{eqnarray}

In the M-step we are trying to find the unique solution of $\phi^{(r+1)}$ such that
\begin{eqnarray}
\frac{\partial}{\partial \phi'} Q(\phi,\phi^{(r)}) \Bigg\vert_{\phi^{(r+1)}} = 0,
\label{Mstep}
\end{eqnarray}
where we define $\phi' = \{\bm{\beta}, (\bm{\mu}_k, \bm{\Sigma}_k), k=1,...,K \}$.
Eq.(\ref{Mstep}) leads to unique solutions \cite{wang2007nonlinear} of
$\bm{\mu}_k^{(r+1)}$, $\bm{\Sigma}_k^{(r+1)}$, $(\sigma^2)^{(r+1)}$.
The updating of $w_k$ can be calculated as the average of the contributions from each subject to the $k$th mixing \cite{wang2007nonlinear}, i.e.,
\begin{eqnarray}
w_k^{(r+1)}
= \frac{1}{n} \sum\limits_{i=1}^{n} \tau_i(k)
=\frac{1}{n}  \displaystyle  \sum\limits_{i=1}^{n} \int g_{ik}(\bm{\theta}_i, \phi^{(r)}) d \bm{\theta}_i .
\label{wk}
\end{eqnarray}

The EM iterates $\phi^{(r)}$ have the important property that the corresponding likelihoods $L(\phi^{(r)})$ are non-decreasing, i.e., $L(\phi^{(r+1)}) \geq L(\phi^{(r)})$ for all $r$ \cite{Wu1983a,wang2007nonlinear}.
\textcolor{black}{
It is also worthwhile to mention that, under certain assumptions,
in SAEM, by splitting the E-step into a simulation-step and a stochastic approximation step \cite{SAEM1999,kuhn2004coupling,KUHN20051020},
or using an improved SAEM which adds a simulation-step before E-step and adds a stochastic approximation step after E-step (MSAEM) \cite{lavielle2014improved}, it is possible to show that the algorithm converges to a local maximum of the likelihood function.
Besides SAEM, there are other algorithms using stochastic methods in order to  converge to a local maximum, see recent references such as \cite{wiens2021robust,brown2021novel,li2021deep}.
}

\subsection{RPEM Algorithm}

From a close look at $g_{ik}(\theta_i, \phi^{(r)})$ in Eq.(\ref{gik}) we find that
\begin{align}
\sum\limits_{k=1}^{K} \int g_{ik}(\bm{\theta}_i, \phi^{(r)}) d \bm{\theta}_i & = 1,
\label{giksum1} \\
\sum\limits_{i=1}^{n} \sum\limits_{k=1}^{K} \int g_{ik}(\bm{\theta}_i, \phi^{(r)}) d \bm{\theta}_i & = n.  \label{giksumn}
\end{align}
Therefore, the unique solutions \cite{wang2007nonlinear} of Eq.(\ref{Mstep}) can be written as,
\begin{align}
\bm{\mu}_k^{(r+1)} &= \frac{ \displaystyle \sum\limits_{i=1}^{n} \int \bm{\theta}_i g_{ik}(\bm{\theta}_i,\phi^{(r)}) d \bm{\theta}_i  }
{ \displaystyle \sum\limits_{i=1}^{n} \int g_{ik}(\bm{\theta}_i,\phi^{(r)}) d \bm{\theta}_i    },
\label{muk}  \\
\bm{\Sigma}_k^{(r+1)} &=  \frac{ \displaystyle\sum\limits_{i=1}^{n} \int ( \bm{\theta}_i - \bm{\mu}_k^{(r+1)} ) ( \bm{\theta}_i - \bm{\mu}_k^{(r+1)} )^T  g_{ik}(\bm{\theta}_i,\phi^{(r)}) d \bm{\theta}_i  }
{ \displaystyle \sum\limits_{i=1}^{n} \int g_{ik}(\bm{\theta}_i,\phi^{(r)}) d \bm{\theta}_i    },
\label{sigmak}  \\
(\sigma^2)^{(r+1)} &= \frac{ \displaystyle \frac{\sum\limits_{i=1}^n m_i}{n}  \left[ \sum\limits_{i=1}^{n} \sum\limits_{k=1}^{K}
\int (\bm{Y}_i -\bm{h}_i(\bm{\theta}_i))^T {\bm{H}_i(\bm{\theta}_i)}^{-1} (\bm{Y}_i -\bm{h}_i(\bm{\theta}_i)   )
 g_{ik}(\bm{\theta}_i,\phi^{(r)}) d \bm{\theta}_i \right]}
 { \displaystyle \sum\limits_{i=1}^{n} \sum\limits_{k=1}^{K} \int g_{ik}(\bm{\theta}_i,\phi^{(r)}) d \bm{\theta}_i  }
 \label{sigma2more} .
\end{align}

In a very useful case \cite{wang2007nonlinear}, we can partition the parameter $\bm{\theta}_i$ into two components:
$\bm{\theta}_i = \{\bm{\alpha}_i, \bm{\beta}_i \}$.
Such that $\bm{\alpha}_i$ is from a mixture of multivariate Gaussians and $\bm{\beta}_i$ is from one single multivariate Gaussian.
In such a case, the EM updates from Eq.(\ref{Mstep}) are given by,
\begin{align}
(\bm{\mu}_{\bm{\alpha}})_k^{(r+1)} &=
\frac{ \displaystyle \sum\limits_{i=1}^n \int \bm{\alpha}_i g_{ik}(\bm{\theta}_i,\phi^{(r)}) d \bm{\theta}_i }
{ \displaystyle \sum\limits_{i=1}^{n} \int g_{ik}(\bm{\theta}_i,\phi^{(r)}) d \bm{\theta}_i },
\label{mualphak}\\
(\bm{\Sigma_\alpha})_k^{(r+1)}  &=
\frac{ \displaystyle \sum\limits_{i=1}^n \int \left[ \bm{\alpha}_i- (\bm{\mu_\alpha})_k^{(r+1)}  \right]
\left[\bm{\alpha}_i- (\bm{\mu_\alpha})_k^{(r+1)} \right]^T g_{ik}(\bm{\theta}_i,\phi^{(r)}) d \bm{\theta}_i }
{ \displaystyle \sum\limits_{i=1}^{n} \int g_{ik}(\bm{\theta}_i,\phi^{(r)}) d \bm{\theta}_i },
\label{sigmaalphak} \\
(\bm{\mu_\beta})^{(r+1)} &=
\frac{ \displaystyle \sum\limits_{i=1}^n \sum\limits_{k=1}^K \int \bm{\beta}_i g_{ik}(\bm{\theta}_i,\phi^{(r)}) d \bm{\theta}_i }
{ \displaystyle \sum\limits_{i=1}^{n} \sum\limits_{k=1}^K \int g_{ik}(\bm{\theta}_i,\phi^{(r)}) d \bm{\theta}_i },
\label{mubeta}  \\
(\bm{\Sigma_\beta})^{(r+1)}  &=
\frac{ \displaystyle \sum\limits_{i=1}^n \sum\limits_{k=1}^K \int \left[ \bm{\beta}_i- (\bm{\mu_\beta})^{(r+1)}  \right]
\left[\bm{\beta}_i- (\bm{\mu_\beta})^{(r+1)} \right]^T g_{ik}(\bm{\theta}_i,\phi^{(r)}) d \bm{\theta}_i }
{ \displaystyle \sum\limits_{i=1}^{n} \sum\limits_{k=1}^K \int g_{ik}(\bm{\theta}_i,\phi^{(r)}) d \bm{\theta}_i }
\label{sigbeta} .
\end{align}

RPEM is a Monte Carlo parametric EM algorithm.
The most important feature of RPEM is that it updates the parameters in Eqs.(\ref{muk}-\ref{sigbeta}) by performing general and efficient Metropolis algorithm \cite{Ceperley95a} based on an `overall' randomized sampling for both discrete labels such as $i$ and $k$ and the continuous variables $\bm{\theta}_i$. Next we describe the basic idea of RPEM in the E-step and M-step.

\subsubsection{E-Step of RPEM}

In order to prepare for the RPEM Monte Carlo integrations in the M-step, we first evaluate the denominator of $g_{ik}(\bm{\theta}_i,\phi)$ which does not depend on $k$. We define it as $N_i$ such that
\begin{align}
N_i =  \sum\limits_{k=1}^{K} w_k n_{ik},
\label{Ni}
\end{align}
where
\begin{align}
n_{ik} \equiv
\int p(\bm{Y}_i\vert\sigma^2, \bm{\theta}_i) p(\bm{\theta}_i\vert\bm{\mu}_k,\bm{\Sigma}_k) d\bm{\theta}_i.
\label{n_ik}
\end{align}

For $n_{ik}$ we sample $\bm{\theta}_i$ from Gaussian $p(\bm{\theta}_i\vert\bm{\mu}_k,\bm{\sigma}_k)$, then we evaluate Eq.(\ref{n_ik}) by taking the average of the samples of $p(\bm{Y}_i\vert\sigma^2, \bm{\theta}_i)$, i.e.,
\begin{align}
n_{ik} \approx \frac{1}{m_{\text{Gauss}}}
\sum\limits_{m=1}^{m_{\text{Gauss}}} p(\bm{Y}_i\vert\sigma^2, \bm{\theta}_i^{(m)})
\Bigg\vert_{\bm{\theta}_i^{(m)} \in p(\bm{\theta}\vert\bm{\mu}_k,\bm{\sigma}_k) }, \label{nikMCave}
\end{align}
where the number of samples $m_{\text{Gauss}}$ is typically set between 200 and 3000 on a single CPU core.
Once we obtain $n_{ik}$ and $N_i$, the $\tau_i(k)$ in Eq.(\ref{tauik}) can be immediately evaluated by
\begin{align}
\tau_i(k) = \frac{w_k n_{ik} }{N_i}.
\end{align}

We also evaluate the log of the likelihood function $L(\phi)$ in Eq.(\ref{L}) as
\begin{align}
\ln L(\phi) = \sum\limits_{i=1}^{n}\ln \left( N_i \right) .
\label{LL}
\end{align}
The program is iterated until $\ln L(\phi)$ stabilizes.

When dealing with models with ODEs, the E-step can be the most time-consuming part.
We typically store the samples of $p(\bm{Y}_i\vert\sigma^2, \bm{\theta}_i) $ calculated from ODEs and obtained for each $i$ and $k$ in Eq.(\ref{nikMCave}), and reuse them in the M-step.

\subsubsection{M-Step of RPEM}

The M-step is important, because it is at the M-step that we truly estimate the parameters.
Inaccurate estimation at each iteration of M-step may cumulate to inaccurate estimations of the parameters and result in unnecessarily long iteration time.

In the M-step we update all the parameters.
The weight in Eq.(\ref{wk}) is calculated from the $n_{ik}$ and $N_i$ obtained in the E-step,
\begin{align}
w_k^{(r+1)} = \frac{1}{n} \sum\limits_{i=1}^{n}
\int g_{ik}(\bm{\theta}_i, \phi^{(r)}) d \bm{\theta}_i
= \frac{w_k}{n}\sum\limits_{i=1}^{n} \frac{n_{ik}}{N_i}.
\label{wkrp1}
\end{align}

The rest of the parameters from Eq.(\ref{muk}) to Eq.(\ref{sigbeta}) can be cast into one type of integral,  which can be generalized as
\footnote{When the integral does not involve looping over the mixture label $k$, we simply treat $K$ in Eq.(\ref{type2int}) as 1.}
,
\begin{align}
\langle f \rangle &=
\frac{ \displaystyle \sum\limits_{i=1}^n \sum\limits_{k=1}^{K} \int f_{ik}(\bm{\theta}_i) g_{ik}(\bm{\theta}_i,\phi^{(r)}) d \bm{\theta}_i }
{\displaystyle \sum\limits_{i=1}^{n} \sum\limits_{k=1}^{K} \int g_{ik}(\bm{\theta}_i,\phi^{(r)}) d \bm{\theta}_i }.
\label{type2int}
\end{align}
Note that,
\begin{align}
\sum\limits_{k=1}^{K} \displaystyle \sum\limits_{i=1}^n \int d \bm{\theta}_i
\left[ \frac{ g_{ik}(\bm{\theta}_i,\phi^{(r)}) }
{\displaystyle \sum\limits_{k=1}^{K} \sum\limits_{i=1}^{n} \int g_{ik}(\bm{\theta}_i,\phi^{(r)}) d \bm{\theta}_i}  \right]
=1,
\label{intgik}
\end{align}
so the integrand of Eq.(\ref{intgik}) can be treated as a target distribution $\pi(s)$,
\begin{align}
\pi(s) =
\frac{ g_{ik}(\bm{\theta}_i,\phi^{(r)})}
{ \displaystyle \sum\limits_{k=1}^{K} \sum\limits_{i=1}^{n} \int g_{ik}(\bm{\theta}_i,\phi^{(r)}) d \bm{\theta}_i}
= \frac{ g_{ik}(\bm{\theta}_i,\phi^{(r)})}{\mathcal{N}}
\label{pistarget}
\end{align}
whose normalization factor is $\mathcal{N}$.

RPEM distinguishes itself from other MCPEM in evaluating Eq.(\ref{type2int}) in the M-step
by directly sampling from $\pi(s)$ in Eq.(\ref{pistarget}).
Because of this,  unlike all the other MCPEM algorithms,
in RPEM we have the advantage of evaluating Eq.(\ref{type2int}) by using unbiased estimator.
As will be shown, RPEM treats continuous variable $\bm{\theta}_i$ and discrete variables $i$ and $k$ at the same footage, based on the profound fact that,
Monte Carlo is a very powerful method in that it applies to not only continuous variables,
but also to discrete variables.
Same techniques have already been widely used in quantum Monte Carlo for several decades  \cite{Lomnitz1981NPA,Ceperley95a,CarlsonRMP15a,CR2020PhD}.

RPEM samples $\pi(s)$ with Metropolis sampling \cite{hammersley2013monte,Ceperley95a}.
We denote $s=\{i, k, \bm{\theta}_i \}$ as the current state, $s'=\{i', k', \bm{\theta}'_{i'} \}$ as the (proposed) new state.
A common choice of the acceptance probability $A(s\rightarrow s')$ is \cite{Kalos1986a},
\begin{equation}
A(s\rightarrow s')=
\min \left[  1, \frac{\pi(s')T(s'\rightarrow s)}{\pi(s) T(s\rightarrow s')} \right] ,
\label{Accptequation}
\end{equation}
where $\pi(s')$ is the new target distribution which equals to $\frac{g_{i'k}(\bm{\theta}'_{i'},\phi^{(r)})}{\mathcal{N}}$.
The proposed transition probability is denoted by $T(s\rightarrow s')$,
which can be further decomposed by
$T(s\rightarrow s') = T(i \rightarrow i') T(k \rightarrow k') T(\bm{\theta}_{i} \rightarrow \bm{\theta}'_{i'})$. The normalization factor $\mathcal{N}$ cancels in Eq.(\ref{Accptequation}).

The Metropolis algorithm is proceeded as follows.
We randomly sample $i'$ and $k'$,
which means both $T(i \rightarrow i')$ and $T(k \rightarrow k')$ are constant, and we sample $\bm{\theta}'_{i'}$ from $p(\bm{\theta}'_{i'}\vert\bm{\mu}_k,\bm{\sigma}_k)$ which means $T(\bm{\theta}_{i} \rightarrow \bm{\theta}'_{i'}) = p(\bm{\theta}'_{i'}\vert\bm{\mu}_k,\bm{\Sigma}_k)$.
After we propose the new state $s'$, the acceptance probability $A(s\rightarrow s')$ in Eq.(\ref{Accptequation}) is concretely written as,
\begin{align}
A(s\rightarrow s') &= \min \left[1,
\frac{g_{i'k'}(\bm{\theta}'_{i'},\phi^{(r)}) \bcancel{T(i' \rightarrow i)} \bcancel{T(k' \rightarrow k)} T(\bm{\theta}'_{i'} \rightarrow \bm{\theta}_{i})  }
{g_{ik}(\bm{\theta}_{i},\phi^{(r)})
\bcancel{T(i \rightarrow i')} \bcancel{T(k \rightarrow k')}
T(\bm{\theta}_{i} \rightarrow \bm{\theta}'_{i'}) }
 \right] \nonumber \\
& = \min \left[1,
\frac{  \frac{ w_{k'}  p(\bm{Y}_{i'}\vert \sigma^2, \bm{\theta}'_{i'}  )
\bcancel{p( \bm{\theta}'_{i'} \vert \mu_{k'}, \Sigma_{k'} )}  }{ N_{i'}  }
\bcancel{p( \bm{\theta}_{i} \vert \mu_k, \Sigma_k )}  }
{ \frac{  w_k p(\bm{Y}_i\vert \sigma^2, \bm{\theta}_{i}  )
\bcancel{p( \bm{\theta}_{i} \vert \mu_k, \Sigma_k )}  }{ N_{i}  }
\bcancel{p( \bm{\theta}_{i'} \vert \mu_{k'}, \Sigma_{k'} )}  }
 \right] \nonumber \\
& = \min \left[1,
\frac{p(\bm{Y}_{i'}\vert \sigma^2, \bm{\theta}'_{i'}  )}
{p(\bm{Y}_i\vert \sigma^2, \bm{\theta}_{i}  ) }
\times \frac{N_i}{N_{i'}} \times \frac{w_{k'}}{w_{k}}
 \right].
\label{MetroAss'2}
\end{align}

We judge
if the new state $s'$ is accepted by Eq.(\ref{MetroAss'2}).
If accepted, we keep the new state $s'$ and set $s'$ as the current state $s$; if not, we keep the current state $s$ and continue to propose new state $s'$.
In fact,
since $N_i$, $N_{i'}$, $w_{k}$ and $w_{k'}$ are obtained by Monte Carlo, there are error bars associate with them.
The Metropolis algorithm of Eq.(\ref{MetroAss'2}) can also take the error bars into account.
We can define the item in Eq.(\ref{MetroAss'2}) as
$\mu_A \equiv \frac{p(\bm{Y}_{i'}\vert \sigma^2, \bm{\theta}'_{i'}  )}
{p(\bm{Y}_i\vert \sigma^2, \bm{\theta}_{i}  ) }
\times \frac{N_i}{N_{i'}} \times \frac{w_{k'}}{w_{k}}$.
Since $N_i$, $N_{i'}$, $w_{k}$ and $w_{k'}$ have error bar, by error propagation,
we can obtain the error bar of $\mu_A$ which can be denote as $\sigma_A$.
So for Eq.(\ref{MetroAss'2}) when we compare if a random number $x$ is smaller than
$\frac{p(\bm{Y}_{i'}\vert \sigma^2, \bm{\theta}'_{i'}  )}
{p(\bm{Y}_i\vert \sigma^2, \bm{\theta}_{i}  ) }
\times \frac{N_i}{N_{i'}} \times \frac{w_{k'}}{w_{k}}$,
we are actually comparing if $x$ is smaller than a Gaussian distributed quantity $A$ whose mean is $\mu_A$ and standard deviation is $\sigma_A$.
We can denote this probability as $P(x<A)$, whose value can be derived as
\begin{align}
P(x<A) = \frac{1+p}{2},
\end{align}
where $p$ is an error function,
\begin{align}
p=erf\left(\frac{\mu_A-x}{\sigma_A \sqrt{2}}\right).
\end{align}
So the Metropolis algorithm in Eq.(\ref{MetroAss'2}), when error bars are considered,
can be described as,
we throw a random number $x$, then we throw another random number $y$,
if $y<P(x<A)$ then $s'$ is accepted and we keep the new state $s'$ and set $s'$ as the current state $s$; if not, keep the current state $s$ and continue to propose new state $s'$.

We continue the Metropolis process until the target distribution is formed and get $m$ sufficient independent samples of $f_{i'k'}^{(j)}(\bm{\theta}_{i'}) $, where $j$ denotes the label of a sample of $(i', k', \bm{\theta}'_{i'})$ combination.
The expectation of Eq.(\ref{type2int}) is evaluated as,
\begin{align}
\langle f \rangle & \approx \frac{1}{m} \sum\limits_{j=1}^{m} f_{i'k'}^{(j)}(\bm{\theta}_{i'})
\bigg\vert_{ \{i', k', \bm{\theta}_{i'}\} \in
\pi(s')
},
\label{expftype2}
\end{align}
and according to central limit theorem \cite{Negele1998Quantum,HammondMCbook},
the error bar of Monte Carlo (also called standard error in statistics) is the standard deviation of the mean of the $m$ independent samples.

The unique Metropolis sampling as shown in Eq.(\ref{MetroAss'2}) makes RPEM not only fast but also accurate,  as will be shown in the following section.

\section{Results and Discussions}\label{secResults}

\subsection{Hardware and Software}\label{sechwsw}

In this paper, we used a ThinkPad P72 laptop with Intel Xeon-2186M CPU (2.9Ghz base frequency and 4.8Ghz max turbo frequency, 6 cores 12 threads) and 64GB DDR4-2666 ECC memory.
RPEM is written in modern Fortran and is fully parallelized using MPI, it can be run on both PCs and supercomputers.
We use Intel Fortran and MPI provided in the free Intel OneAPI 2022.1.3.
For the comparison among RPEM, SAEM, and QRPEM, all the runs are using 6 cores and the platform is Windows 10 Pro for Workstation
\footnote{
We also tested the speed of RPEM using GFortran and MPICH on Linux and Mac with M1 chip,
the speed are consistent and similar with using Intel OneAPI.
}.
The Monte Carlo sampling size ${m_{\text{Gauss}}}$ in Eq.(\ref{nikMCave}) is set to 1000 for RPEM. In RPEM, we use FLINT \cite{FLINT} which is written in modern Fortran as our non-stiff ODE solver, and when stiff ODEs are detected, we use the latest version of the well-known LSODA solver in ODEPACK \cite{hindmarsh1983odepack}.
The absolute tolerance ATOL and relative tolerance RTOL in the ODE solvers are set to $10^{-6}$ in RPEM.
SAEM is in Monolix 2019R1 with default settings.
QRPEM is in Certara RsNLME \cite{RsNLME} with NLME Engine version 21.11.2, and its sample size is set to 500, and the maximum ODE steps is set to 500.

\subsection{Model with analytic solution}\label{secanalmodel}
We first report the results of RPEM
by using the same two-stage model as in \cite{wang2007nonlinear} with the useful case as indicated by Eqs.(\ref{mualphak}-\ref{sigbeta}).
It is a one-compartment PK model with $K=2$ mixing whose plasma concentration is given by
\begin{equation}
y_{ji} = \frac{D}{V_i} e^{-k_i t_j} \left(1+ \epsilon_{ji}\right),
\label{Yji}
\end{equation}
where $j$ ranges from 1 to the number of samples per subject, $m_i$, and $i$ ranges from 1 to the number of subjects, $n$.
The random effects are denoted by $\epsilon_{ji}$, which is a Gaussian random number with standard deviation $\sigma$.  $D$ is a bolus drug administration with the unit of dose, $V_i$ is the volume and $k_i$ is the elimination rate constant.

At the first stage, $\bm{Y}_i \sim N(\bm{h}_i(\bm{\theta}_i),\sigma^2 \bm{H}_i(\bm{\theta}_i))$, for this model it explicitly means,
\begin{align}
\bm{h}_i(\bm{\theta}_i) &= \frac{D}{V_i} \left( e^{-k_i t_1}, ...,  e^{-k_i t_{m_i}}   \right)^T, \\
\bm{H}_i(\bm{\theta}_i) &= \frac{D^2}{V^2_i} \text{diag} \left( e^{-2 k_i t_1}, ...,  e^{-2 k_i t_{m_i}}   \right),
\end{align}
and therefore,
\begin{align}
p(\bm{Y}_i\vert\sigma^2, \bm{\theta}_i) &=
\prod_{j=1}^{m_i}
\frac{    \exp \left\{  -\frac{1}{2} \left[\frac{Y_{ji} - \frac{D}{V_i}\exp(-k_i  t_j)}{\sigma \left\vert \frac{ D}{V_i}\exp(-k_i t_j) \right\vert }\right]^2   \right\}      }
{ \sigma  \left\vert\frac{D}{V_i}\exp(-k_i t_j) \right\vert  \sqrt{2\pi}} .
\end{align}

At the second stage, we have
\begin{align}
\bm{\theta}_i=(k_i,V_i)^T \sim_{i.i.d} \sum\limits_{k=1}^{K} w_k N(\bm{\mu}_k, \bm{\Sigma}_k),
\end{align}
where $\bm{\mu}_k=(\mu_k,\mu_V)^T$ and we assume $\bm{\Sigma}_k = \text{diag}(\sigma_k^2, \sigma_V^2)$. Therefore,
\begin{align}
p(\bm{\theta}_i\vert\bm{\mu}_k,\bm{\sigma}_k) &=
\frac{\exp \left[ -\frac{1}{2} \left( \frac{k_i-\mu_k}{\sigma_k} \right)^2 \right] }{\sigma_k \sqrt{2\pi}}
\times
\frac{\exp\left[ -\frac{1}{2}\left( \frac{V_i-\mu_V}{\sigma_V} \right)^2 \right]}{\sigma_V \sqrt{2\pi}}.
\end{align}

For the $K=2$ case used in this paper, $\bm{\mu}_k$ and $\bm{\Sigma}_k$ can be concretely written as $\bm{\mu}_1 = (\mu_{k1}, \mu_V)^T$, $\bm{\mu}_2 = (\mu_{k2}, \mu_V)^T$, $\bm{\Sigma}_1 = \text{diag}(\sigma_{k1}^2, \sigma_V^2)$ and
$\bm{\Sigma}_2 = \text{diag}(\sigma_{k2}^2, \sigma_V^2)$.
We fix $D=100$, and we set $m_i=5$ such that $t_1$ to $t_5$ are 1.5, 2, 3, 4, 5.5 correspondingly.
All the parameters we used to generate the data $\bm{Y}_i$
\footnote{
Observation data $y_{ji}$ for analytic model can be generated from Eq.(\ref{Yji}).
}
are listed in the ``True Values'' column in Table \ref{Table1}.
The initial condition of the parameters is listed in the ``Initial Values'' column.
Starting from some initial condition, our purpose is to use RPEM to find the true values of the parameters.

\begin{table}[!hptb]
\caption{
Parameter values estimated by RPEM for varying subject numbers, $n$, each using 50 iterations. True values are simulated, and initial conditions for each experiment are indicated. Time to completion is included in the column headers. Data are presented as the mean.
The value in parentheses is the $\sigma$ on that digit, e.g., 19.93(5) means $19.93 \pm 0.05$,
which, is the error bar (standard deviation of the mean) of the independent samples \cite{Negele1998Quantum} after convergence.
For example, if we run 50 iterations, and after 20 iterations convergence was reached, then we take the uncorrelated samples from the last 30 iterations, and the $\sigma$ is the standard deviation of the mean of those uncorrelated samples \cite{HammondMCbook,martin2016interacting}.
}
\label{Table1}%
\begin{tabularx}  {\textwidth}{cc*4{>{\centering\arraybackslash}X}c}
\toprule & True Values   & \begin{tabular}{@{}c@{}} $n=100$ \\ 0.4 sec \end{tabular} & \begin{tabular}{@{}c@{}} $n=2\times 10^4$ \\ 55 sec \end{tabular} & \begin{tabular}{@{}c@{}} $n=2\times 10^6$ \\ 72 min \end{tabular} & \begin{tabular}{@{}c@{}} $n=4\times 10^7$ \\ 30 core-hrs \end{tabular}
& Initial Values \\ \hline
$\mu_V$         & 20    & 19.93(5)  & 20.006(8) & 20.008(5)  & 20.011(4)  & 50 \\ \hline
$\mu_{k1}$      & 0.3   & 0.301(1)  & 0.3005(3) & 0.2994(3)  & 0.3000(3)  & 1.0  \\ \hline
$\mu_{k2}$      & 0.6   & 0.599(2)  & 0.5973(4) & 0.5994(5)  & 0.5990(5)  & 1.0  \\ \hline
$w_1$           & 0.8   & 0.8016(6) & 0.7992(4) & 0.7996(5)  & 0.7985(6)  & 0.5\\ \hline
$w_2$           & 0.2   & 0.1983(6) & 0.2007(4) & 0.2004(5)  & 0.2015(6)  & 0.5  \\ \hline
$\sigma_{V}$    & 2     & 2.01(2)   & 1.97(1)   & 1.967(5)   & 1.966(2)   &  50/3 \\ \hline
$\sigma_{k1}$   & 0.06  & 0.057(1)  & 0.0592(1) & 0.0593(2)  & 0.0597(2)  & 1/3 \\ \hline
$\sigma_{k2}$   & 0.06  & 0.048(1)  & 0.0602(7) & 0.0610(4)  & 0.0615(6)  & 1/3 \\ \hline
$\sigma$        & 0.1   & 0.102(1)  & 0.1000(1) & 0.10036(2) & 0.10027(2) & 0.3\\
\bottomrule
\end{tabularx}
\end{table}

\begin{figure}[!hptb]
\centering
\includegraphics[width=\columnwidth]{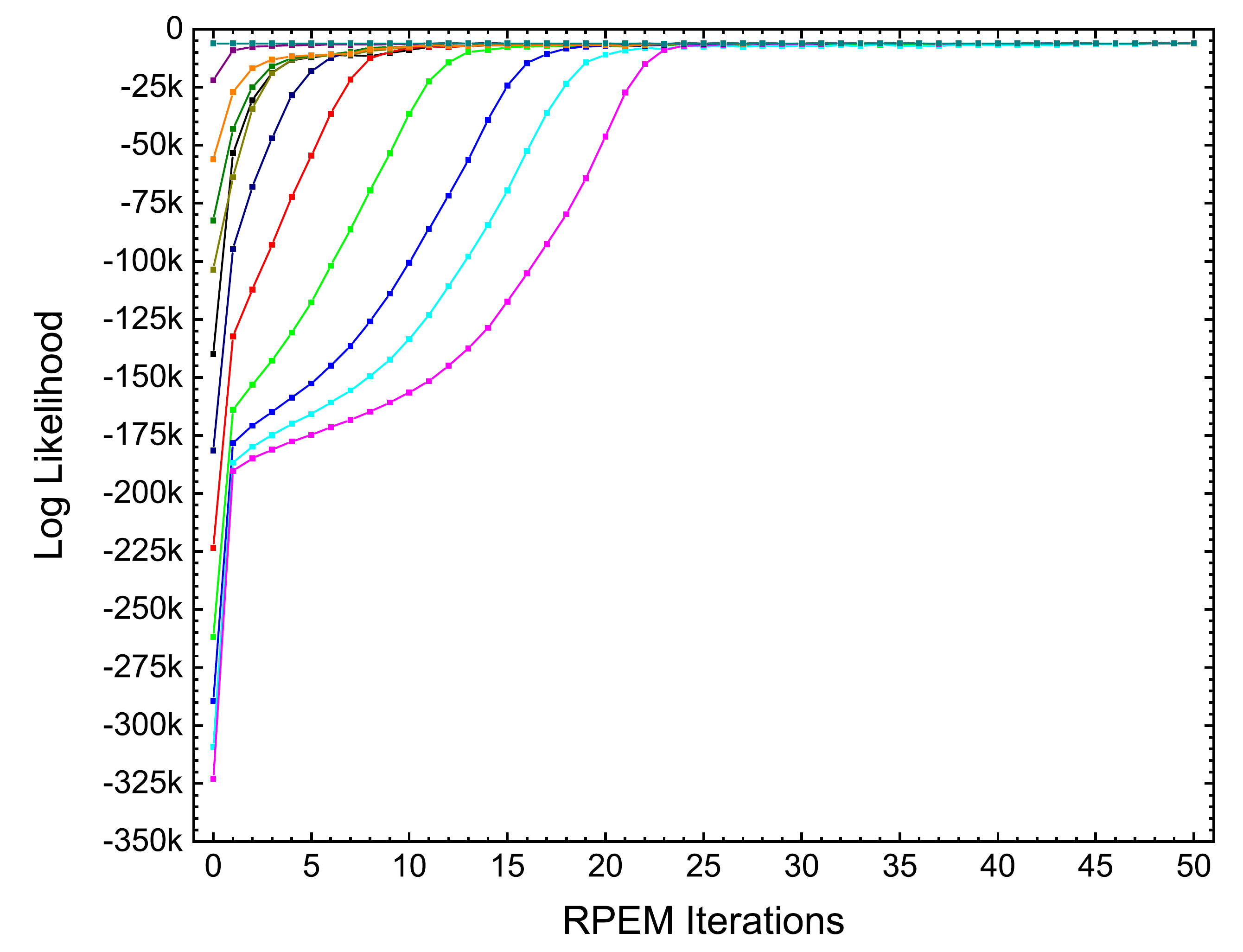}
\caption{The convergence of RPEM's log-likelihood starting from 11 initial conditions plus starting from the true values. }
\label{convergence}
\end{figure}

In Table \ref{Table1}, we report the evaluated parameters using RPEM for 50 iterations.
\textcolor{black}{RPEM is fast and accurate even on a single CPU core, as can be seen in the $n=100$, $n = 2\times 10^4$ and $n = 2 \times 10^6$ cases.
For $n=100$ case with 50 iterations, RPEM took 0.4 seconds with results comparable to those obtained by tradition MCPEM \cite{wang2007nonlinear} using importance sampling which took about half an hour in 2007 using Matlab.
Because RPEM is fast and accurate, it enables us to deal with large datasets efficiently.
Unlike generating 200 data sets each with $n=100$ (so  $n=2\times 10^4$ total), running MCPEM on each of the 200 datasets individually, and finally combining the results as was done in \cite{wang2007nonlinear} and  took approximately 100 hours, in RPEM we finished running for the whole $n=2\times 10^4$ subjects in 55 seconds with even more accurate results.
Furthermore, we finished the running for 2 million subjects in only 72 minutes and the results remained almost the same.}

Note that in RPEM, we can flexibly adjust the number of Gaussian samples in the E-step and the number of Metropolis samples in the M-step; therefore, RPEM does not have a strong subject number $n$ dependence.
For example, for $n=100$ it took 0.4 second, but for $n=2\times 10^4$ which is 200 times bigger than 100 it only took 55 seconds instead of $0.4\times 200 =80 $ seconds. Similarly, $n=2 \times 10^6$ is bigger than $n=2\times 10^4$ by a factor of 100, but RPEM only took 72 minutes instead of $55/60 \times 100=91.7$ minutes.
In all the training, we kept the number of Gaussian samplings in the E-step as $2000$,  the number of Metropolis trials in the M-Step ranged from $2n$ to $200n$, and the autocorrelation time \cite{martin2016interacting} was about 80 steps.

For $n=4\times 10^7$ (i.e., the population of the whole of California) we used MPI and 5 CPU cores. Since the performance of each CPU core on a laptop will change depending on how many cores are fully activated, the measurement of the speedup of MPI on a laptop is not exactly accurate. Nonetheless, RPEM still was able to finish the training with 5 cores in a very reasonable 6.1 hours (30.5 core hours).

RPEM is not only fast and accurate, but also robust.
We extensively tested it with varying initial conditions, and we consistently converged within the first 50 iterations.
In Fig. \ref{convergence}, we show the convergence of the log-likelihood of RPEM for the case of $n=2 \times 10^4$ starting from 11 different initial conditions. The log-likelihoods are calculated in the E-step using Eq.(\ref{LL}).
We particularly picked various poor initial conditions whose starting log-likelihood was as low as $-3.23\times 10^5$,
and for all of them, RPEM was able to rapidly reach the stabilized log-likelihood which was around $-6000$. For better initial conditions RPEM was able to reach stabilized log-likelihood within 15 iterations. For worse initial conditions RPEM was able to reach stabilized log-likelihood within the first 30 iterations.
The almost horizontal line on the top with dark cyan color is RPEM's evolution of log-likelihood starting from the true values.

\subsection{Model with ordinary differential equations}\label{secodemodel}

In this section,
our task is to use RPEM, SAEM and QRPEM to reconstruct the $\bm{\mu}_k$ and $\bm{\Sigma}_k$ from the simulated data
based on a realistic Voriconazole model \cite{Neely2015a} with ordinary differential equations (ODEs), and compare RPEM with SAEM and QRPEM in terms of speed and accuracy.

\subsubsection{Voriconazole Model}

In this Voriconazole model, we follow the data and model format for Pmetrics \cite{neely2012Pmetrics,yamada2020npag}.
The 7 primary parameters are $K_a$, $V_{max0}$, $K_m$, $V_{c0}$, $F_{A1}$, $K_{cp}$, and $K_{pc}$.
So we have,
\begin{equation}\label{thetaiode}
\bm{\theta}_i = \left(K_a,V_{max0},K_m,V_{c0},F_{A1},K_{cp},K_{pc}\right)^T,
\end{equation}
and the population mean is
\begin{equation}\label{mukode}
\bm{\mu}_k =
\left(\mu_{K_a},\mu_{V_{max0}},\mu_{K_m},\mu_{V_{c0}},\mu_{F_{A1}},\mu_{K_{cp}},\mu_{K_{pc}}\right)^T,
\end{equation}
and again we assume $\bm{\Sigma}_k
= \text{diag}(\sigma_{K_a}^2, \sigma_{V_{max0}}^2, \sigma_{K_m}^2, \sigma_{V_{c0}}^2, \sigma_{F_{A1}}^2, \sigma_{K_{cp}}^2, \sigma_{K_{pc}}^2)$.
The covariate is weight ($wt$).
The secondary parameters which are obtained from primary parameters and covariate are $V_m$ and $V$,
\begin{align}
V_m &= V_{max0} \times {wt}^{0.75},
\\
V &= V_{c0} \times {wt}.
\end{align}

For any subject $i$, ODEs are listed as below,
\begin{align}
\frac{d x_1}{dt} &= -K_a \times x_1,
\\
\frac{d x_2}{dt} &= -K_a \times x_1
+ r^{(i)}_{\rm{IV}}(t)
- \frac{V_m^{(i)} \times x_2}{K_m \times V^{(i)} + x_2}
-K_{cp} \times x_2
+K_{pc} \times x_3,
\\
\frac{d x_3}{dt} &= K_{cp} \times x_2
-K_{pc} \times x_3,
\end{align}
where for subject $i$, $r^{(i)}_{\rm{IV}}(t)$ is the ratio between dose and duration at time $t$. If at time $t$ the dose is non-zero and duration is zero, it means a bolus and $x_1(t)$ needs to be added by an additional $\textrm{dose} \times F_{A1}$. $V_m^{(i)}$ and $V^{(i)}$ are its secondary parameters $V_m$ and $V$.

The concentration for subject $i$ at time $t_j$ is given by,
\begin{equation}\label{yjiode}
y_{ji} = \frac{x_2(t_j)}{V^{(i)}} + \epsilon_{ji},
\end{equation}
and we assume the noise $\epsilon_{ji}$ is a Gaussian random number whose standard deviation $\sigma_{ji}$ has the following form,
\begin{align}
\sigma_{ji} = c_0
+ c_1 \times \frac{x_2(t_j)}{V}
+ c_2 \times \left[\frac{x_2(t_j)}{V}\right]^2
+ c_3 \times \left[\frac{x_2(t_j)}{V}\right]^3 ,
\label{sigmaji}
\end{align}
where $c_0$ to $c_3$ are non-negative constants.

Similar with what has been described in Sec. \ref{secanalmodel},
at the first stage, $\bm{Y}_i \sim N(\bm{h}_i(\bm{\theta}_i), \bm{H}_i(\bm{\theta}_i))$, and for this model we have
\begin{align}
\bm{h}_i(\bm{\theta}_i) &= \frac{1}{V^{(i)}} \left[ x_2(t_1), ...,  x_2(t_{m_i}) \right]^T, \\
\bm{H}_i(\bm{\theta}_i) &= \text{diag} \left( \sigma^2_{1i}, ..., \sigma^2_{m_ii}   \right),
\end{align}
and therefore,
\begin{align}
p(\bm{Y}_i\vert\sigma_{ji}^2, \bm{\theta}_i) &=
\prod_{j=1}^{m_i}
\frac{    \exp \left\{  -\frac{1}{2} \left[\frac{Y_{ji} - \frac{x_2(t_j)}{V^{(i)}}}{\sigma_{ji} }\right]^2   \right\}      }
{ \sqrt{2\pi} \sigma_{ji}  } .
\label{Yjiode}
\end{align}
At the second stage, again we have
$\bm{\theta}_i \sim_{i.i.d} \sum\limits_{k=1}^{K} w_k N(\bm{\mu}_k, \bm{\Sigma}_k)$.

\subsubsection{Simulated Data}

For the simulated data,
we set the number of subjects $n=50$.
We assume $K=1$, the covariate is weight.
For each subject, we set observation time $t$ (unit is hour) as 2, 4, 6, 8, ..., 48 so the number of observations is $m_i=24$.
At $t=0$ we set dose as 180 and duration time as 2.
At $t=24$ we set dose as 180 and duration time as 0.
Similarly, with what was done in \cite{liu2016comparing},
the covariate weights are all set to 16.5 (unit is kg), to minimize the effects/noise from covariates when reconstructing the population parameters $\bm{\mu}_k$ and $\bm{\Sigma}_k$.
For the noise in Eq.(\ref{sigmaji}), we set $c_0=0.02$, $c_1=0.1$, and $c_2=c_3=0$.

\begin{figure}[!hptb]
\centering
\includegraphics[width=\columnwidth]{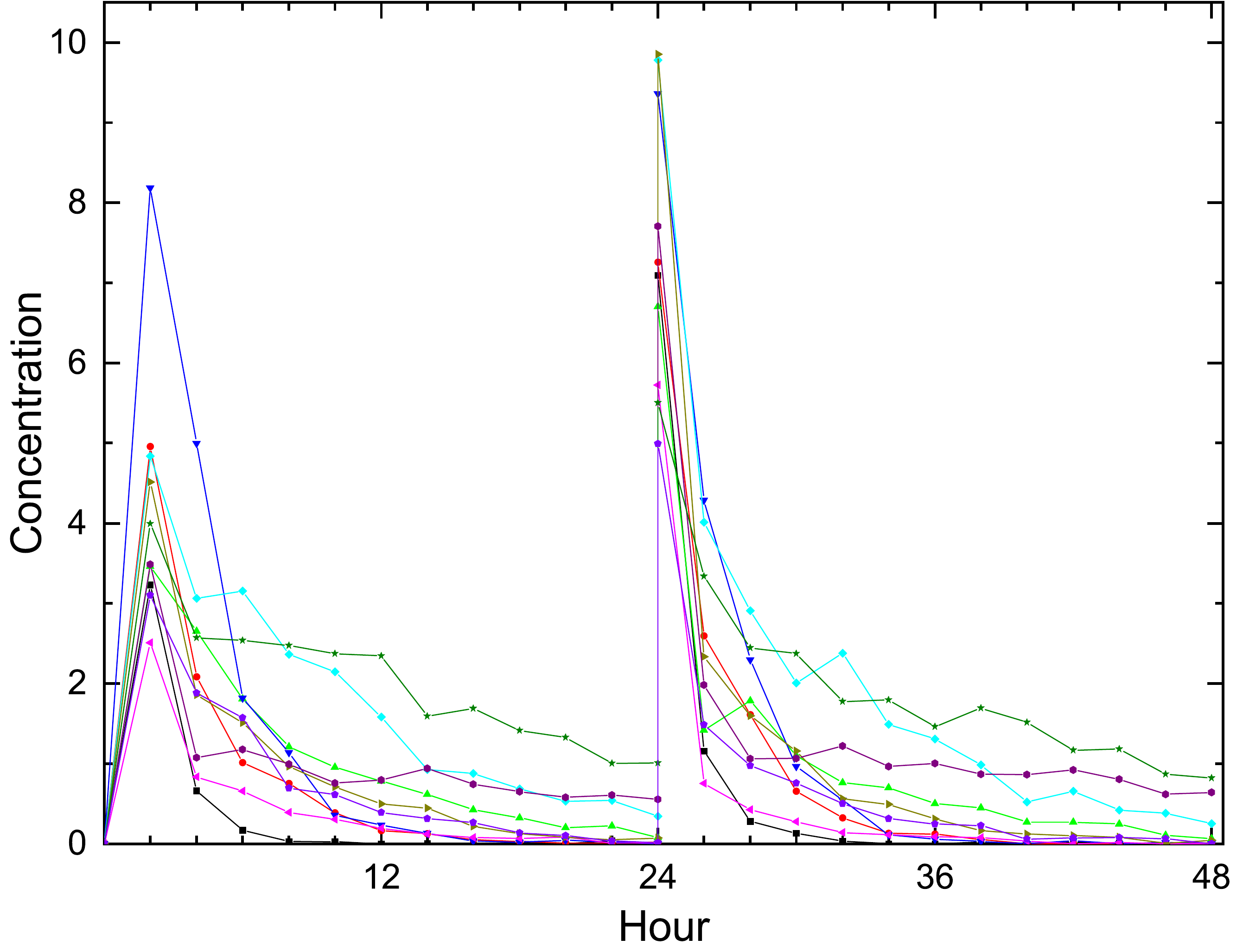}
\caption{Illustration of the simulated concentration vs. time. 10 randomly selected subjects are used. }
\label{Vori_showall}
\end{figure}

The simulated data file for the Voriconazole ODE model can be found in \cite{Voridatafix}.
The observation concentration data $y_{ji}$ is provided as the ``OUT'' column. They can be generated by using the 50 sets of primary parameters in \cite{Vorisimparfix}, which are directly sampled from Gaussian whose true values of $\bm{\mu}_k$ and $\bm{\Sigma}_k$ are listed as the "True" rows in Table \ref{tabcomp}. No log transformations of the primary parameters are needed for our data.
We randomly select 10 subjects and illustrate the simulated concentration vs. time in Fig. \ref{Vori_showall}.

\subsubsection{Stopping Criterion}
RPEM's stopping criterion is described as follows.
As the iteration is going on, we took the latest consequent 30 runs' log likelihood, calculate their slope based on least square method \cite{llsq}.
Before convergence, this slope must be positive.
After convergence, the log likelihood is flattened, and this slope cannot be always positive anymore, it can be positive or negative.
Once we detected at which iteration such a negative slope first occurs, we stopped the iteration, and we take the samples of the latest consequent 30 runs as stabilized samples.

None of the stabilized samples are wasted.
We re-sample from these stabilized samples using Eq.(\ref{MetroAss'2}) and obtain our final estimations of $\bm{\mu}_k$ and $\bm{\Sigma}_k$.
Besides, since the stabilized samples of $\bm{\theta}_i$ in RPEM are approximately distributed from $\sum\limits_{k=1}^{K} w_k N(\bm{\mu}_k, \bm{\Sigma}_k)$,
we also included a fast Gaussian mixture clustering algorithm \cite{FinklerGMM} which directly find the optimum $\bm{\mu}_k$ and $\bm{\Sigma}_k$ from the stabilized samples of $\bm{\theta}_i$ as well. Such results are marked as RPEM-GMM.
So, when RPEM finishes, it will have RPEM-GMM results also.
When report the speed of RPEM, the cost of RPEM-GMM is already included.

\subsubsection{Comparison among RPEM, SAEM, and QRPEM}

We randomly picked 21 initial conditions from \cite{Vorisimparfix} which is used to generate the simulated data.
The 21 initial conditions are picked from id number
1,
10,
13,
15,
17,
20,
23,
25,
27,
3,
30,
33,
35,
37,
40,
43,
45,
47,
5,
50,
7,
by setting the corresponding primary parameters as the initial $\bm{\mu}_k$, and $\bm{\Sigma}_k=\bm{\mu}_k/2.5$.
We let RPEM, SAEM and QRPEM each starts from such 21 initial conditions and compare their speed and accuracy in reconstructing true population parameters $\bm{\mu}_k$ and $\bm{\Sigma}_k$.
We use $\sigma_{K_a}, \sigma_{V_{max0}}, \sigma_{K_m}, \sigma_{V_{c0}}, \sigma_{F_{A1}}, \sigma_{K_{cp}}, \sigma_{K_{pc}}$ to represent $\bm{\Sigma}_k$.

\begin{figure}[!hptb]
\centering
\includegraphics[width=\columnwidth]{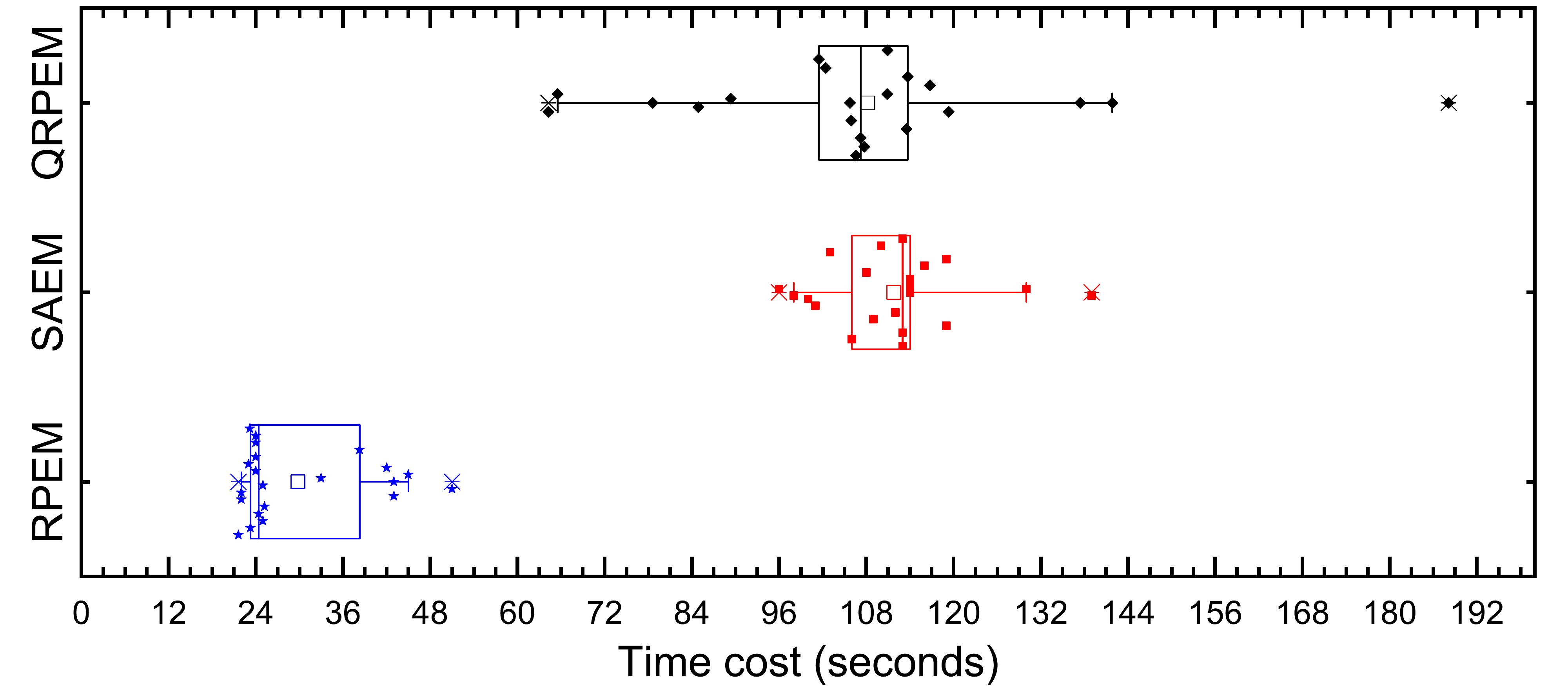}
\caption{The speed comparison among RPEM, SAEM and QRPEM for the Voriconazole model.
The timing of the runs is represented as data by solid symbols.
We use box plot plus data overlap.
The definition of the box plot is as follows,
the whiskers denote the 5\% and 95\% percentile of the data,
the body of the box covers 25\% to 75\% percentile,
the mean and median are represented by a hollow square and a vertical line correspondingly. }
\label{time_cost}
\end{figure}

\begin{figure} [b!]
  \centering
  \includegraphics[width=\textwidth]{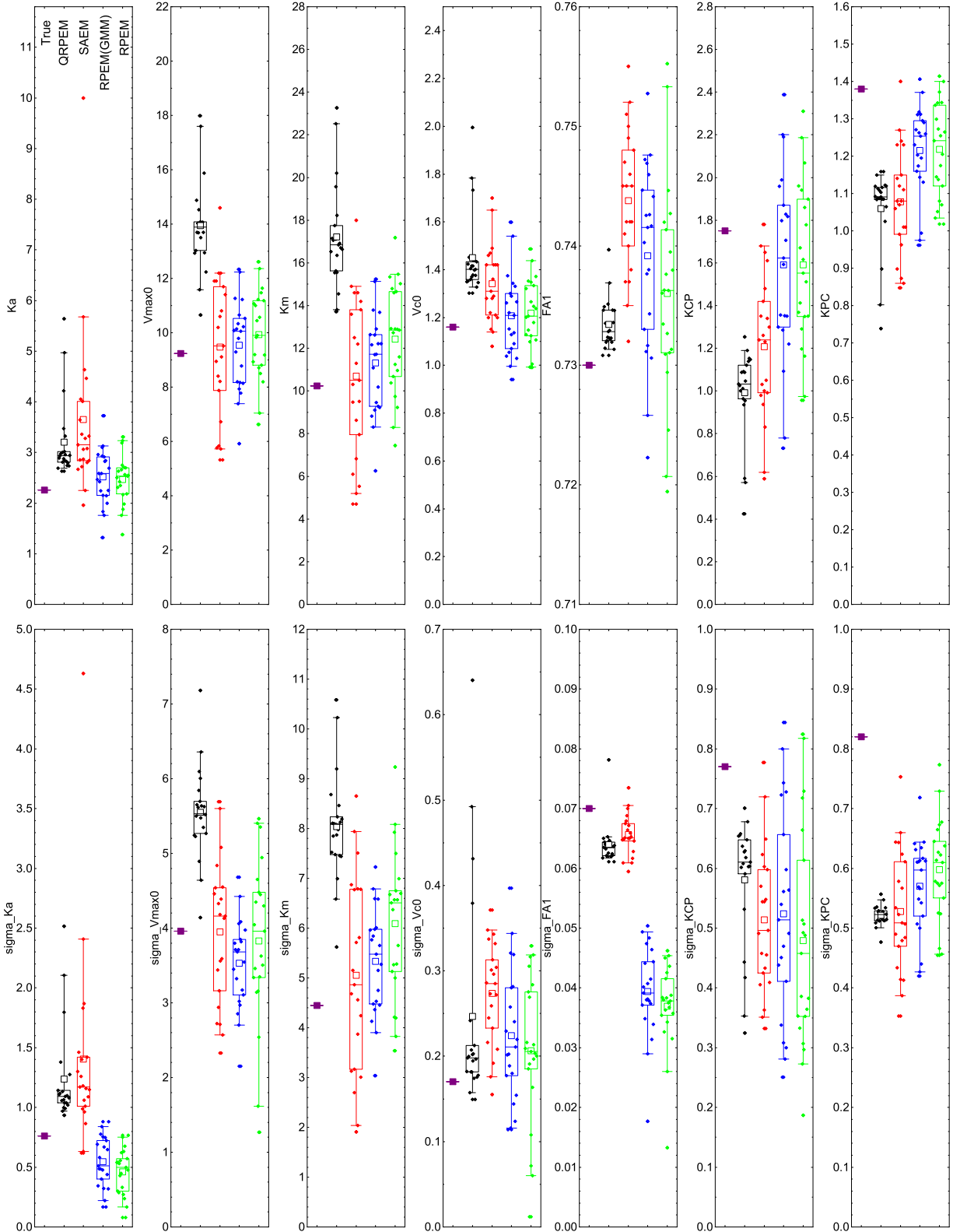}
  \caption{(Caption next page.)}
  \label{parameters2575}
\end{figure}
\addtocounter{figure}{-1}
\begin{figure} [t!]
  \caption{(Previous page.) The comparison among QRPEM, SAEM, and RPEM in terms of the ability of reconstructing $\bm{\mu}_k$ and $\bm{\Sigma}_k$.
From left to right, the leftmost symbol is true value,
the black box with data overlap are QRPEM's result,
the red box with data overlap are SAEM's result,
the blue ones are RPEM-GMM's result,
the rightmost green ones are RPEM's result.
The definition of the box with data overlap is the same as in Fig. \ref{time_cost}. }
\end{figure}

\begin{figure}[!hptb]
\centering
\includegraphics[width=\columnwidth]{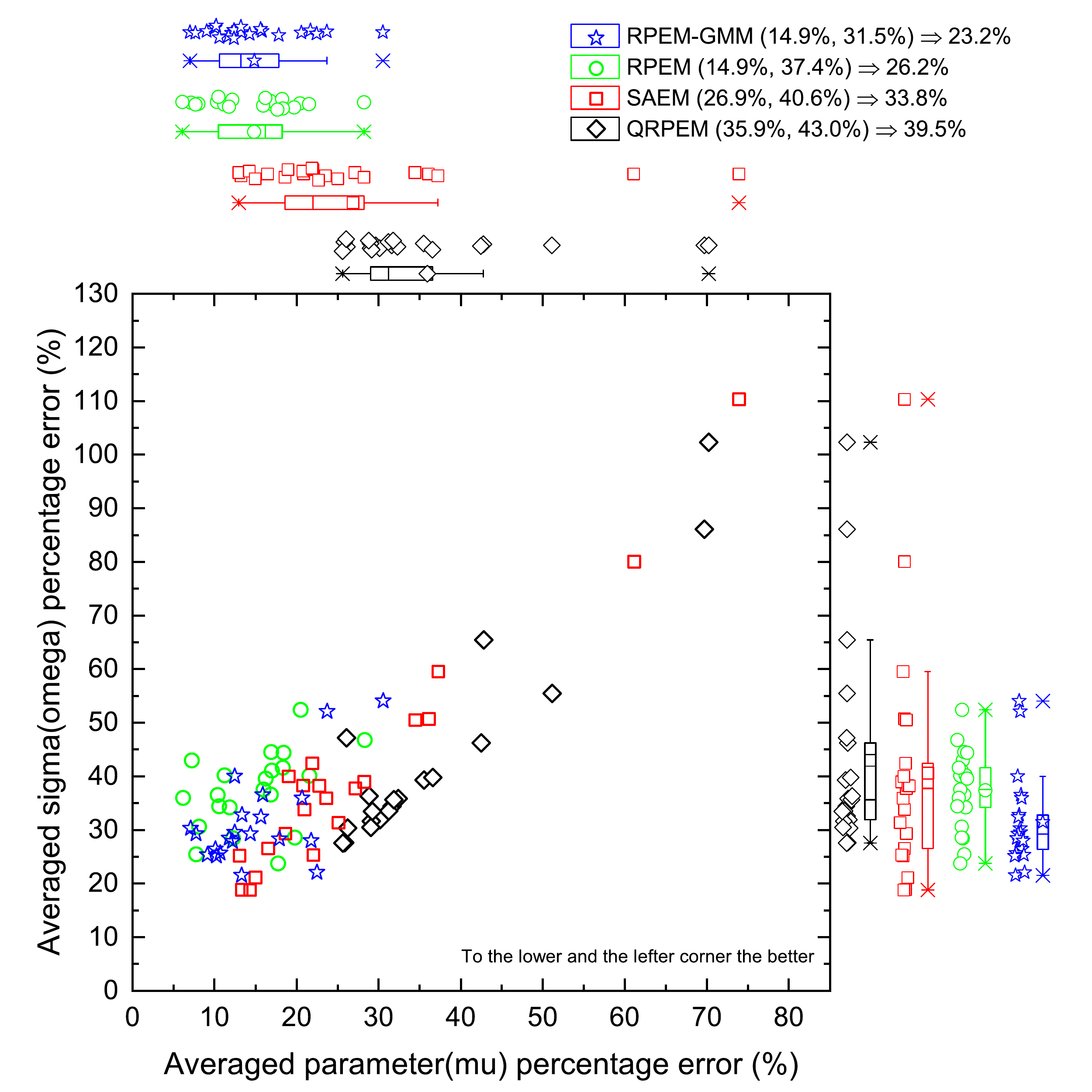}
\caption{The comparison among RPEM, SAEM and QRPEM in terms of averaged percentage error per sigma (also called omega in SAEM) .vs. averaged percentage error per parameter.
The red squares are from SAEM,
the green circles are from RPEM,
and the blue stars are from RPEM-GMM.
The box plots with data overlap on the top and right side have the same definition as in Fig. \ref{time_cost}.  }
\label{ptg_error}
\end{figure}

\begin{table}[!htbp]
\caption{The parameter reconstruction comparisons among RPEM, SAEM, and QRPEM.
The results are obtained by averaging the results from the 21 randomly picked initial conditions.
The overall averaged percentage errors for all the $\bm{\mu}_k$ and $\bm{\Sigma}_k$ are listed in the parentheses correspondingly.
} \label{tabcomp}
\begin{tabularx}  {\textwidth}{l*7{>{\centering\arraybackslash}X}}
\toprule
Method \&. \% error/parameter & $\mu_{K_a}$        & $\mu_{V_{max0}}$    & $\mu_{K_m}$        & $\mu_{V_{c0}}$       & $\mu_{F_{A1}}$         & $\mu_{K_{cp}}$        & $\mu_{K_{pc}}$        \\
\midrule
True & 2.26 & 9.23 & 10.32  & 1.16 & 0.73 & 1.75 & 1.38 \\
RPEM(14.9\%) & 2.463  & 9.928  & 12.423  & 1.218 & 0.736 & 1.591 & 1.218 \\
RPEM-GMM(14.9\%) & 2.516  & 9.543  & 11.321  & 1.208 & 0.739 & 1.592 & 1.215 \\
SAEM(26.9\%) & 3.651	& 9.472	& 10.686 &	1.341 &	0.744	& 1.208 &	1.078	\\
QRPEM(35.9\%) & 3.204	& 13.943	& 17.223 &	1.450 &	0.733	& 0.991 &	1.059	\\
\toprule
Method \&. \% error/sigma & $\sigma_{K_a}$        & $\sigma_{V_{max0}}$    & $\sigma_{K_m}$        & $\sigma_{V_{c0}}$       & $\sigma_{F_{A1}}$      & $\sigma_{K_{cp}}$        & $\sigma_{K_{pc}}$
\\
\midrule
True & 0.76 & 3.96 & 4.45  & 0.17 & 0.07 & 0.77 & 0.82 \\
RPEM(37.4\%) & 0.462  & 3.829  & 6.093  & 0.206 & 0.0369 & 0.479 & 0.598 \\
RPEM-GMM(31.5\%) & 0.547  & 3.528  & 5.330  & 0.224 & 0.0394 & 0.524 & 0.570 \\
SAEM(40.6\%) & 1.403	& 3.951	& 5.055	& 0.274	& 0.0656	& 0.514	& 0.527 \\
QRPEM(43.0\%) & 1.235	& 5.548	& 8.036	& 0.246	& 0.0639	& 0.580	& 0.521 \\
\bottomrule
\end{tabularx}
\end{table}

We list the speed comparison among RPEM, SAEM and QRPEM in Fig. \ref{time_cost}.
We see that most RPEM runs finished at around 20 to 25 seconds (about 42 iterations, each RPEM iteration takes about 0.6 seconds).
The rest of RPEM runs finished between 30 to 60 seconds.
For SAEM which are represented by the red symbols,
most of them finished between 90 and 120 seconds,
and few of them finished between 130 and 144 seconds.
The results of QRPEM are more scattered,
the fast runs finish around 64 seconds,
while the slow ones take around 192 seconds.
On average, we find that for Voriconazole model,
the speed of SAEM and QRPEM are comparable (QRPEM is slightly faster),
and RPEM is about 3 or 4 times faster than both of them.

In Fig. \ref{parameters2575},
we show RPEM, SAEM, and RPEM's abilities in reconstructing the true population parameters $\bm{\mu}_k$ and $\bm{\Sigma}_k$.
We find that almost all the true $\bm{\mu}_k$ and $\bm{\Sigma}_k$ are covered by RPEM within $5-95\%$ percentile.
Out of the total 14 population parameters,
RPEM only missed 2 of them, namely $\sigma_{F_{A1}}$ and $\sigma_{K_{pc}}$ are not covered within $5-95\%$ percentile.
SAEM on the other hand, missed 7 out of 14,
$\mu_{F_{A1}}$,
$\mu_{K_{cp}}$ and $\sigma_{K_{cp}}$,
$\mu_{K_{pc}}$ and $\sigma_{K_{pc}}$,
$\mu_{K_a}$ and $\sigma_{V_{c0}}$.
Within $25-75\%$ percentile,
RPEM only missed 6 out of 14,
$\sigma_{V_{c0}}$,
$\sigma_{F_{A1}}$ and $\sigma_{F_{A1}}$,
$\sigma_{K_{cp}}$, $\mu_{K_{pc}}$ and $\sigma_{K_{pc}}$.
While SAEM missed 10 out of 14,
namely
$\mu_{K_a}$ and $\sigma_{K_a}$,
$\mu_{V_{c0}}$ and $\sigma_{V_{c0}}$,
$\mu_{F_{A1}}$ and $\sigma_{F_{A1}}$,
$\mu_{K_{cp}}$ and $\sigma_{K_{cp}}$,
$\mu_{K_{pc}}$ and $\sigma_{K_{pc}}$.
For QRPEM,
we find it nearly missed all the all the true $\bm{\mu}_k$ and $\bm{\Sigma}_k$.
However, we find some ODE difficulties in QRPEM due to the additive parameterizations for the Voriconazole model,
therefore, we suspect its accuracy is compromised because of that.

To quantitatively compare the accuracy among RPEM, QRPEM, and SAEM,
in Fig. \ref{ptg_error},
for all of them, we plot the averaged percentage error for all the $\bm{\Sigma}_k$ .vs.
averaged percentage error for all the $\bm{\mu}_k$, for each of the 21 runs \footnote{
In each run,
each of the $\sigma_{K_a}$, $\sigma_{V_{max0}}$, $\sigma_{K_m}$, $\sigma_{V_{c0}}$, $\sigma_{F_{A1}}$, $\sigma_{K_{cp}}$, and $\sigma_{K_{pc}}$ has a corresponding percentage error compared with their true values.
We sum up these 7 percentage errors and divide it by 7,
and that is how we obtain the averaged percentage error for all the $\bm{\Sigma}_k$.
Averaged percentage error for all the $\bm{\mu}_k$ is similarly obtained.
}.
The symbols are to the lower and the left corner the better.
We found that both RPEM and RPEM-GMM are more concentrated towards the lower and the left corner
than SAEM and QRPEM.
Both SAEM and QRPEM's averaged percentage error for all the $\bm{\mu}_k$ are bigger than RPEM's.
The overall averaged percentage error for all the $\bm{\mu}_k$
\footnote{
The overall averaged percentage error for all the $\bm{\mu}_k$ means that, we averaged the 21 runs' averaged percentage error for all the $\bm{\mu}_k$.
Same thing applies to the overall averaged percentage error for all the $\bm{\Sigma}_k$.
}
for both RPEM and RPEM-GMM are about $14.9\%$, for SAEM is $26.9\%$, for QRPEM is $35.9\%$.
The overall averaged percentage error for all the $\bm{\Sigma}_k$ for RPEM and RPEM-GMM are about $37.4\%$ and $31.5\%$, for SAEM is $40.6\%$, for QRPEM is $43.0\%$.
We also notice that SAEM and QRPEM's outliers can be about two times of RPEM's. 									
In Table \ref{tabcomp},
we list the averaged $\bm{\mu}_k$ and $\bm{\Sigma}_k$ reconstructed from the 21 initial conditions of RPEM, RPEM-GMM, SAEM, and QRPEM.

Based on the analysis of Fig. \ref{time_cost}, Fig. \ref{parameters2575}, Fig. \ref{ptg_error},
as well as Table \ref{tabcomp}, we can qualitatively conclude that, for the Voriconazole model, RPEM is faster and more accurate than SAEM and QRPEM.

We also need to point out that, for RPEM, by using the Metropolis algorithm Eq.(\ref{MetroAss'2}),
we did not neglect any of the 50 subjects in any iterations in any of the RPEM runs.
All 50 subjects are always almost equally sampled,
i.e., in the M-step in each iteration in each RPEM run,
each of the 50 subjects shares about $2\%$ of all the samples in Eq.(\ref{expftype2}).

\subsubsection{Scalability}

Since RPEM is equipped with MPI, we also run RPEM from an initial condition on Agave supercomputer cluster at Arizona State University to test its scalability.

\begin{figure}[!hptb]
\centering
\includegraphics[width=\columnwidth]{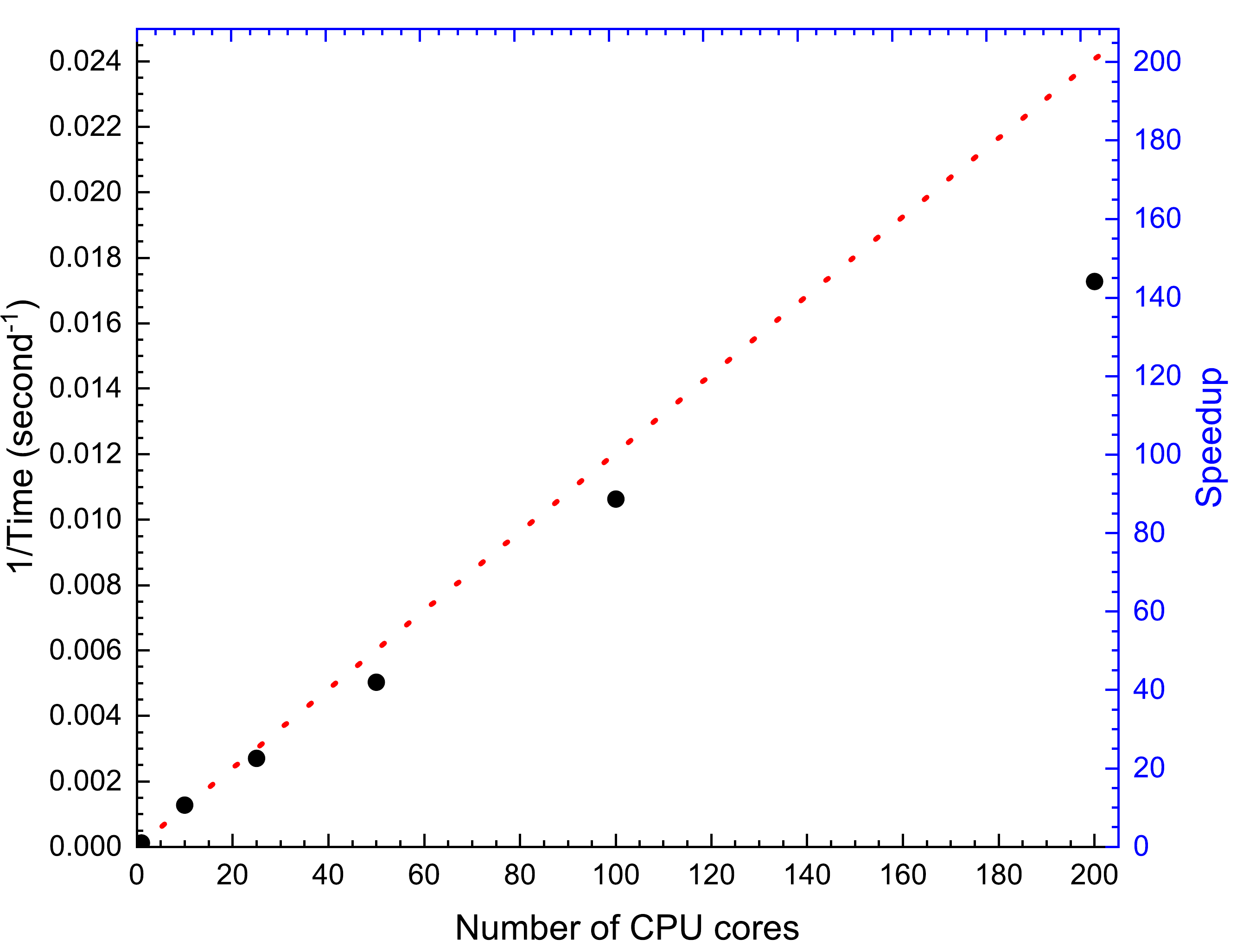}
\caption{The scaling efficiency of the current RPEM code on Agave supercomputer cluster at Arizona State University. We use $m_{\textrm{Gauss}}=3\times 10^4$ in the E-step, and $10^8$ trials ($m \approx 2.5 \times 10^5$ uncorrelated samples) in the Metropolis judgement in M-step. The red dotted line is the theoretical speedup. The back dots are real speedup.}
\label{speedup}
\end{figure}

In Fig. \ref{speedup}, we show the scaling efficiency of RPEM.
Overall, when the number of CPU cores are less than 100 (which covers the range from a laptop to high-end personal desktop nowadays), the efficiency of RPEM is around $90\%$.
As the number of cores beyond 100, since the samples at E-step needs to be calculated on each core become fewer and fewer, the real computation time in solving ODEs is decreased, so the percentage of MPI communication time in the total time increased.
Therefore the efficiency decreased to about $70\%$ when the number of CPU cores beyond 200.
However, when the ODE model is complicated enough and require many samples, such that as long as we are able to make computation time always much more than MPI communication time, RPEM on supercomputer will always have reasonably high efficiency.
So, RPEM is ready for models which are complicated enough and no longer be suitable on a personal computer.

\section{Summary and Outlook}
\label{secSummary}

In this paper,
we presented our quantum Monte Carlo inspired novel MCPEM algorithm which we call RPEM.
RPEM distinguishes itself from other MCPEM algorithms mostly in the M-step,
which uses Metropolis algorithm and samples both continuous variables and discrete variables at the same time efficiently.
Therefore, unlike other MCPEM algorithms which uses biased estimators,
both RPEM's E-step and the M-step uses unbiased estimators which will lead to fast convergence and accurate results.

With concrete examples of a one-compartment two-mixture analytic model and a Voriconazole model with ODEs,
and by comparing RPEM with SAEM and QRPEM,
we show that RPEM is indeed not only a fast, but also an accurate MCPEM algorithm.
We also show that RPEM is a scalable high performance MCPEM which can be run on supercomputers for more complicated models.

In future work we will further test and validate RPEM on more complex data and models, test several approaches to rapidly arrive at the optimal number of mixing components, and
further develop or implement techniques to avoid local maxima.

We wish as a newly developed MCPEM method, RPEM can be a useful addition to the current MCPEM methods.
We welcome ideas, suggestions, and cooperating opportunities from the community.

\section*{Acknowledgments}
R.C. thanks Professor Kevin E. Schmidt at Arizona State University for inspiring discussions and insights.
R.C. thanks Dr. Keith Nieforth at Certara, Inc. for the immense help in preparing the Voriconazole model and data files for Certara RsNLME, which are used in the paper in benchmarking the QRPEM results.
R.C. thanks Dr. Michael Tomashevskiy at Certara, Inc. for the suggestions in setting the QRPEM engine parameters.
R.C. acknowledge Research Computing at Arizona State University for providing HPC and storage resources that have contributed to the research results reported within this paper.
R.C also acknowledge valuable help received from Fortran-lang community \cite{curcic2021toward,sof2022}. In particular, R.C. thanks Ondřej Čertík, John Campbell, Milan Curcic, Martin Diehl, Steve Kargl, Steve Lionel (Doctor Fortran), Bharat Mahajan, Vincent Magnin, Arjen Markus, Panagiotis Papasotiriou, Ivan Pribec,  Vivek Rao, Brad Richardson, Simon Rowe, Amir Shahmoradi, Michal Szymanski, Theodore-Thomas Tsikas, John S. Urban, and Yi Zhang for various and generous help in modern Fortran coding and the ODE solvers used in RPEM code.

\section*{Declarations}
This work was funded in part by U01 1FD006549 (Neely, PI).
This work has no conflict of interest/competing interests.
%\bmhead{Data availability}

\bibliographystyle{apsrev4-2}
%\bibliography{../../../Bibliography/CRBIB}
%apsrev4-2.bst 2019-01-14 (MD) hand-edited version of apsrev4-1.bst
%Control: key (0)
%Control: author (72) initials jnrlst
%Control: editor formatted (1) identically to author
%Control: production of article title (-1) disabled
%Control: page (0) single
%Control: year (1) truncated
%Control: production of eprint (0) enabled
%apsrev4-2.bst 2019-01-14 (MD) hand-edited version of apsrev4-1.bst
%Control: key (0)
%Control: author (8) initials jnrlst
%Control: editor formatted (1) identically to author
%Control: production of article title (0) allowed
%Control: page (0) single
%Control: year (1) truncated
%Control: production of eprint (0) enabled
%

\end{document}